# Jahn–Teller-like Distortion in a One-dimensional π-Conjugated Polymer

Ziyi Wang[1,3,4†], Boyu Qie[2,4,5†], Weichen Tang[1,3†], Jingwei Jiang[1,3], Fujia Liu[2], Peter H. Jacobse[1], Jiaming Lu[1], Xinheng Li[1], Steven G. Louie[1,3*], Felix R. Fischer[2,3,4,5*], Michael F. Crommie[1,3,4*]

[1] Department of Physics, University of California, Berkeley, CA 94720, USA.
[2] Department of Chemistry, University of California, Berkeley, CA 94720, USA.
[3] Materials Sciences Division, Lawrence Berkeley National Laboratory, Berkeley, CA 94720, USA.
[4] Kavli Energy NanoSciences Institute at the University of California Berkeley and the Lawrence Berkeley National Laboratory, Berkeley, CA 94720, USA.
[5] Bakar Institute of Digital Materials for the Planet, Division of Computing, Data Science, and Society, University of California, Berkeley, CA 94720, USA.
[*] Correspondence to: crommie@berkeley.edu, ffischer@berkeley.edu, sglouie@berkeley.edu
[†] These authors contributed equally to this work.

**Abstract:**

Structurally distorting low-dimensional π-conjugated systems can profoundly influence their electronic properties, but controlling such behavior in extended-width systems remains challenging. Here we demonstrate that a one-dimensional conjugated polymer—*poly*-(difluorenoheptalene-ethynylene) (PDFHE)—undergoes a pronounced out-of-plane backbone distortion, equivalent to a spontaneous symmetry breaking (SSB) of its mirror symmetry. We synthesized PDFHE on noble metal surfaces and characterized its structure and electronic states using low-temperature scanning tunneling microscopy. Rather than adopting a planar, high-symmetry conformation, PDFHE relaxes into non-planar isomers stabilized by a Jahn–Teller-like mechanism that relieves an electronic instability relative to the gapped planar structure. Density functional theory calculations corroborate these findings, revealing that distortion lowers the total polymer energy and enlarges the bandgap, providing a microscopic explanation for the SSB. Our results show that even in mechanically robust extended π-systems, subtle electron-lattice coupling can spontaneously drive significant structural rearrangements.

## INTRODUCTION

Spontaneous symmetry breaking (SSB), the emergence of asymmetry from a high-symmetry state without external intervention, is a fundamental concept underpinning many important phenomena in condensed matter physics, such as crystallization,[1, 2] magnetism,[3] charge density waves[4-6] and superconductivity.[7, 8] In molecular solids and nanostructures, SSB favors structural distortion when the energetic cost of maintaining high-symmetry is outweighed by the stabilization gained from adopting a lower-symmetry conformation. Such structural distortion can have profound consequences, stabilizing new phases and electronic states. Classic examples include the Jahn–Teller distortion in fullerene anions[9-11] and octahedral transition metal complexes,[12, 13] bond-length alternations arising from antiaromaticity in cyclic carbon rings,[14-16] and Peierls bond dimerization in polyacetylene chains.[17, 18]

Achieving symmetry-breaking distortions in extended one-dimensional (1D) and two-dimensional (2D) π-conjugated systems, however, remains a significant challenge. As the dimensionality of (or width of) a quasi-1D π-system increases, mechanical rigidity and multiple bonding pathways tend to preserve higher-symmetry configurations, making it difficult to induce SSB-related structural distortions.[19] For example, Peierls-like distortions in metallic graphene nanoribbons and extended polyacene-based polymers are typically suppressed by the increased rigidity of the system stabilizing high-symmetry conformations.[20-22] Significant effort has been spent performing on-surface synthesis of extended π-conjugated polymer chains, yet the resulting polymers have largely retained high-symmetry planar geometries in their ground state without exhibiting spontaneous lattice distortion.[22-27]

Here we demonstrate that finite-width 1D conjugated polymers can undergo Jahn–Teller-like distortions in their ground state. We show that *poly*-(difluorenoheptalene-ethynylene) (PDFHE), a polymer featuring an extended nonalternant π-system, relaxes into a non-planar geometry as a direct result of electronic structure-driven SSB. Low-temperature scanning tunneling microscopy (STM) reveals a periodic out-of-plane deformation along the polymer backbone, providing the first real-space observation of a Jahn–Teller-like distortion in a robust covalent polymer chain. The observation of a sizable electronic bandgap in PDFHE measured by scanning tunneling spectroscopy (STS) is supported by DFT simulations. This work demonstrates that even mechanically rigid, extended π-conjugated polymers can undergo SSB, thus creating new opportunities for the design of nanomaterials that are responsive to intrinsic electronic/structural coupling.

## RESULTS/DISCUSSION

### Design of a Jahn–Teller-like distortion

To induce structural distortion in a 1D conjugated π-system it is essential to create a large enough energy difference between highly symmetric (e.g., planar) and distorted conformations to overcome any mechanical restoring force. Our strategy to accomplish this involved introducing non-benzenoid (i.e., non-hexagonal) rings into the π-system, thus imparting strain by deviating from an ideal trigonal planar (120°) bonding geometry and disrupting the conjugation typical of fully benzenoid systems.[28, 29] This produces molecules with partial open-shell (radical) character that makes them more susceptible to symmetry-lowering distortions.[30-35] We identified difluorenoheptalene (DFH) as an ideal building block for constructing a conjugated π-system with diverse ring structure (Figure 1A). DFH (a structural isomer of bisanthene) is a polycyclic aromatic hydrocarbon (PAH) consisting of a fused framework of four 6-membered rings, two 7-membered rings, and two 5-membered rings. Recent studies have shown that this non-benzenoid PAH possesses approximately 72% diradical character in its singlet ground state[36] (Figure 1A), originating from the synergistic effects of ring strain (induced by the presence of 5- and 7-membered rings) and the loss of conventional benzenoid resonance stabilization.

To access an extended polymer we chose an ethynylene (–C≡C–) linker group to bridge the 7,14-positions between DFH cores (Figure 1A). The carbon–carbon triple bond of the ethynylene provides a rigid, linear, conjugated linker while minimizing steric hindrance between neighboring DFH units (i.e., ensuring that hydrogens on adjacent cores do not sterically interfere). The bonding along the backbone of ethynylene-linked DFH polymers (denoted PDFHE) can be described by contributions from two resonance forms having very different electronic configurations as shown in Figure 1B. In the first (left side), each DFH unit adopts an open-shell diradical state whereas the linker is described by a carbon–carbon triple bond (leaving two unpaired electrons on each DFH core). In the second (right side) the unpaired electrons are incorporated into a cumulene-like double bond (=C=C=) between DFH units, yielding an overall closed-shell configuration for the polymer. Simple localization considerations suggest that the carbon–carbon triple bond and associated unpaired electrons should be energetically unfavorable compared to the more extended cumulene linkage. Close inspection of the cumulene structure, however, implies that it leads to bond-angle strain because the interior angle (marked θ) of the carbon atom bonded to the alkynyl linker would be larger than the ideal trigonal planar bond angle (120°) since it is the interior angle of a heptagon (ideally 128.6°). This strain can be relieved by rotating the carbon atom out of the molecular plane, thus causing the PDFHE polymer to bend (i.e., by hinging the two bonds connected to the 6-membered rings (Figure 1C)).

This intuitive picture of SSB is supported by density functional theory (DFT) calculations of the ground-state energy of PDFHE at different distortions. As shown in Figure 1C, a strained planar structure lies at a high-energy saddle point in the simulated configuration space, whereas symmetry-broken distorted structures are observed at much lower energies. The lowest energy state is the all-*cis*-PDFHE (by –883 meV/unit cell relative to the strained planar structure). The all-*trans*-PDFHE conformation lies at a slightly higher energy than all-*cis*-PDFHE, but is still significantly lower in energy than the planar structure (by –651 meV/unit cell). The origin of this Jahn–Teller-like spontaneous symmetry breaking is the energetically costly polyradical state, a situation that is reminiscent of the degenerate manifolds seen for conventional Jahn–Teller precursors.[37]

**On-surface Synthesis of PDFHE**

The synthesis of 7,14-bis(dichloromethylene)-7,14-dihydroheptaleno[2,1,10,9-jklm:4,5,6,7-j'k'l'm']difluorene (**1**), the molecular precursor for PDFHE is depicted in Figure 2A. A Ramirez reaction of heptalenodifluorendione **2** with $CCl_4$ installs geminal dichloromethylene groups (=$CCl_2$) at either end of the heptalene core.[38] Thermal activation of these strategically positioned =$CCl_2$ groups on Au(111) substrates followed by reductive coupling of the intermediate carbon-centered radicals induces on-surface polymerization and gives rise to ethynylene linkers in PDFHE. We performed on-surface polymerization by depositing precursor **1** onto a clean Au(111) surface under UHV via thermal evaporation from a Knudsen cell, achieving sub-monolayer coverage. After deposition the molecules spontaneously self-assembled on the gold surface as seen in the STM image of Fig. 2B. Here a densely packed island of **1** on Au(111) can been seen that exhibits a honeycomb-like packing motif. The higher-resolution STM image of Fig. 2C reveals that each precursor molecule has a distinct boat-shaped consisting of two bright lobes attributed to the dichloromethylene groups that are bridged by a dimmer central region (the DFH core). This appearance is consistent with the expected geometry of **1** and is reproduced by DFT simulations of isolated precursors (see Supporting Information Fig. S1A).

Analytically pure samples of **1** were deposited onto a clean Au(111) surface under ultrahigh vacuum (UHV) from a Knudsen cell evaporator, achieving sub-monolayer coverage. Figure 2B shows a representative topographic STM image of densely packed islands of precursor **1** self-assembled into honeycomb-like packing motif driven by intermolecular van der Waals (vdW) interactions. Higher-resolution STM images (Figure 2C) reveal that on Au(111) each precursor molecule adopts a saddle-like conformation (Figure S1A). Two bright lobes protruding from the surface can be assigned to the $=CCl_2$ groups lining either end of the central DFH core, the dimmer region in topographic images (see molecular model overlay in Figure 2C). This preferred adsorption geometry is consistent with the lowest energy conformation of **1** predicted by DFT simulations of isolated precursors in the gas phase (Figure S1A).

The step-growth polymerization of **1** was induced by heating molecule-decorated surfaces to $T = 230$ °C for 2 h. Topographic STM images recorded following the annealing process exclusively show covalently linked 1D-chains of PDFHE that span tens of nanometers, occasionally even exceeding 100 nm in length (Figure 2D). A close-up STM image reveals that PDFHE features a highly periodic segmented structure with remarkably few defects (Figure 2E). PDFHE chains are often aligned with the herringbone reconstruction of the Au(111) substrate, suggesting a preferred adsorption along the *fcc* domains. Bond-resolved STM (BRSTM) imaging with CO-functionalized tips (Figure 2F) shows the internal bonding of the 1D-polymer, including the DFH cores and the ethynylene linkers. Similar experiments performed on Ag(111) highlight the robustness of the PDFHE growth from molecular precursors (Figure S2). Because Ag(111) has no pronounced surface reconstruction, PDFHE chains align into densely packed parallel arrays driven by non-covalent interchain vdW interactions.

**Structural Distortion in PDFHE**

Closer investigation of BRSTM images recorded on PDFHE chains on Au(111) reveal a periodic modulation of the brightness contrast along the polymer chain that is suggestive of a *z*-height corrugation along the polymer backbone (Figure 2F). This feature is even more pronounced in high-resolution constant-current STM topographic images recorded on individual polymer segments (Figure 3A). Here the DFH cores exhibit a distinct trapezoidal shape and are tilted out of the *xy*-plane (thus breaking two of the internal mirror symmetries of the unit cell, see Supporting Information Note S1). A 3D projection of the BRSTM image of Figure 2F shown in Figure 3B further highlights the periodicity and topography of the corrugation. As seen in Figure 3C, a *z*-height profile recorded along a line-cut (red arrow in Figure 3A) shows an apparent *z*-height modulation of $\Delta z = 17 \pm 3$ pm with a period of $\lambda = 0.95 \pm 0.10$ nm, consistent with the all-trans conformation of PDFHE theoretically predicted in Figure 1C (polymers grown on Ag(111) similarly exhibit an all-trans conformation, see Figure S3). Curiously, neither on Au(111) nor on Ag(111) substrates were we able to observe PDFHE chains adopting the all-*cis* conformation predicted by DFT simulation to be 232 meV lower in energy that the all-*trans* conformation (Figure 1C). We attribute this discrepancy between theory and experiment to stabilizing contributions arising from the interaction of the polymer with the metal growth substrate, which were not included in our calculation. Further evidence of substrate interaction is provided by the fact that the apparent *z*-height modulation of all-trans polymers on Au(111) is only ~17 pm (Figure 3C) rather than the 66 pm predicted based on DFT simulations (Figure 3D).

Although we did not observe all-*cis* PDFHE chains in our STM experiments, we can identify isolated *cis*-defects within otherwise all-*trans* polymers (see Figure S4A). These defects are relatively rare and occur in only ~10% of the PDFHE polymers. In STM topographic maps *cis*-defects appear as a single DFH unit with rectangular symmetry that is flanked on either side by units that tilt in opposite directions (Figure S4B,C). Both the density and structure of *cis*-defects for PDFHE polymers grown on Au(111) (Figure S4) and Ag(111) (Figure S3) are comparable, suggesting a similar surface interaction that favors the all-*trans* conformation.

**Electronic Structure of all-*trans* PDFHE**

Having determined the geometric distortion induced by SSB in all-*trans* PDFHE chains, we set out to probe their electronic structure using STS. Figure 4A shows representative $dI/dV$ point spectra acquired at different positions on an all-*trans*-PDFHE polymer chain marked by colored crosses on the BRSTM image inset ($dI/dV$ spectra reflect the local density of states (LDOS) under the STM tip). Irrespective of position, all spectra exhibit a sharp rise in conductance at $V_S \approx +1.0$ V (*feature* 1), indicating the onset of an unoccupied electronic state. Spectra recorded along the edge of DFH cores (red curve) show a second pronounced onset at $V_S \approx -0.5$ V (*feature* 2), which corresponds to an occupied state. The notable absence of spectral features across a wide bias range $-0.5$ V < $V_S$ < +1 V suggests that *feature* 1 can be assigned to the conduction band (CB) edge of the polymer and, correspondingly, *feature* 2 represent the position of the valence band (VB) edge. From the energy separation of *features* 1 and 2 we estimate an electronic bandgap of 1.5 ± 0.3 eV for all-*trans*-PDFHE on Au(111).

Differential conductance maps recorded along the backbone of an all-*trans*-PDFHE show vanishing LDOS within the band gap, a typical feature for semiconductors. $dI/dV$ maps at sample biases corresponding to *features* 1 and 2 instead show distinct nodal patterns. A differential conductance map recorded at $V_S = +0.96$ V (*feature* 1, Figure 4B) shows a mirror-symmetric distribution in the wavefunction (with respect to the polymer main axis) and an antinode at the center of each DFH unit. The wavefunction distribution corresponding to the filled-state VB edge recorded at $V_S = -0.48$ V (*feature* 2, Figure 4D), by contrast, is characterized by a central node-like feature bisecting the DFH core. An apparent gradient in the LDOS intensity and nodal patterns follows the tilt within each DFH unit along the polymer backbone.

To corroborate our experimental results and gain further insight into the electronic structure emerging from a Jahn–Teller-like distortion, we performed *ab initio* DFT-PBE simulations for a freestanding all-*trans* PDFHE polymer at the theoretically determined structure. The experimental band gap of 1.5 ± 0.3 eV derived from STS is reasonably consistent with a theoretical predicted indirect bandgap of $E_g$ ~1.0 eV (Figure 4F,G and Figure S5A). DFT Kohn-Sham eigenvalues with the PBE approximation for the exchange-correlation potential are known to underestimate semiconducting band gaps and in this study we do not account for interaction effects with the underlying substrate.[39, 40] Figures 4C,E show the theoretical LDOS maps corresponding to the lower CB and the upper VB edge, respectively. The CB edge map features an antinode along the center of the polymer backbone (Figure 4C) whereas the VB wavefunction exhibits a node at the same position (Figure 4E), in reasonable agreement with the corresponding d$I$/d$V$ maps (Figures 4B,D). This agreement supports our assignment of *feature* 1 as the CB minimum and *feature* 2 as the VB maximum of all-*trans*-PDFHE.

A deeper understanding of the Jahn–Teller-like SSB in PDFHE polymers can be gained through DFT calculations of their band structures across three distinct conformations. When we alter the structure from the high-symmetry planar configuration to the all-*trans* and then to the all-*cis* conformation, we observe a progressive opening of the bandgap: ~0.9 eV in the planar form, increasing to ~1.0 eV for all-*trans*, and then to ~1.1 eV for the all-*cis* conformation (Figure S5). This trend reflects a decrease in total energy, indicating enhanced electronic stability, thus supporting our hypothesis that structural distortions in PDFHE are driven by a Jahn–Teller-like mechanism. By lowering the molecular symmetry the polymer alleviates electronic instabilities and gains a more favorable electronic structure characterized by a larger energy gap and reduced total energy.

## CONCLUSIONS

In conclusion, we observe spontaneous symmetry breaking in a 1D π-conjugated polymer synthesized by linking open-shell molecular building blocks with ethynylene linkers. The resulting PDFHE polymer exhibits a Jahn–Teller-like distortion of its backbone, adopting an all-*trans* conformation on both Au(111) and Ag(111) surfaces to access a lower-energy closed-shell ground state. PDFHE polymers exhibit the characteristics of a robust semiconducting band structure with an apparent indirect bandgap of ~1.5 eV, reasonably consistent with DFT calculations. The presence of two low-energy closed-shell conformations (*trans* vs. *cis*) for the polymer backbone suggests that, in principle, external stimuli may be used to reversibly switch the polymer between two bistable states. This offers a tantalizing route toward future nanoscale actuators or molecular machines that exploit molecular symmetry breaking.

## EXPERIMENTAL SECTION

**Precursor Synthesis.** Full details of the synthesis and characterization of **1** are provided in the Supporting Information.

**UHV Growth and Scanning Probe Experiments.** Atomically-clean Au(111) and Ag(111) substrates were prepared using alternating $Ar^+$ sputtering and annealing cycles under UHV conditions. Sub-monolayer coverages of precursor **1** were attained through UHV deposition using a homebuilt Knudsen cell evaporator operating at a crucible temperature of 180 °C for 20 min with the substrate held at room temperature. Polymerization was induced by gradually raising the substrate temperature (~5 °C / min) up to $T$ = 230 °C and held for 120 min to induce the growth of PDFHA.

Each stage of growth was characterized by imaging the surface with a commercial Createc low-temperature STM operated at $T$ = 4.7 K using a chemically etched tungsten tip. STS measurements were conducted using a lock-in amplifier operated with a modulation frequency $f$ = 533 Hz and a wiggle voltage $V_{AC}$ = 4 mV. d$I$/d$V$ maps were conducted in constant-height mode with a CO-functionalized tip and a wiggle voltage $V_{AC}$ = 30 mV. BRSTM images were obtained by mapping the d$I$/d$V$ signal at low-bias ($V_S$ = 0 mV) using constant-height scans. STS peak positions were determined by fitting them with Lorentzian distributions. Parameters for each state being based on averaging ~10 spectra from each of 3 polymers using different STM tips. STM images were processed using WSxM software.[41]

**Calculations.** Simulations for different structural configurations of PDFHE: First-principles density functional theory (DFT) calculations were per-formed using the PBE exchange-correlation functional, as implemented in the Quantum ESPRESSO package.[42, 43] Norm-conserving (NC) pseudopotentials were employed,[44] with an energy cutoff of 100 Ry and a Gaussian smearing of 0.002 Ry. To ensure the accuracy of the results, a vacuum layer of 10 Å was applied along all non-periodic directions. The atomic structures were fully relaxed until the forces on all atoms were smaller than 0.01 eV/Å. Band structure calculations were performed using 40 $k$-points along the periodic direction.

Potential energy diagram of PDFHA structural configurations: The planar, *cis*, and *trans* structures were first fully relaxed to their respective equilibrium configurations. Structures with fractional contributions from two different configurations were generated via linear interpolation between the corresponding atomic positions. For example, let $\boldsymbol{r}_i^{plane}$ and $\boldsymbol{r}_i^{cis}$denote the positions of the $i$-th atom in the planar and *cis* structures, respectively. A structure composed of a fraction $p$ of the planar and $1-p$ of the *cis* configuration has atomic position given by

$$\boldsymbol{r}_i^{p\ plane;(1-p)\ cis} = p\,\boldsymbol{r}_i^{plane} + (1-p)\,\boldsymbol{r}_i^{cis}$$

The total energy of each interpolated structure was obtained through DFT calculations using the same computational parameters described above.

Simulation on ground state configuration of precursor **1**: First principles DFT calculations were performed using the Gaussian 16 software package.[45, 46] Gas phase geometry optimization was performed using the B3LYP exchange–correlation functional with the def2-SVP basis set.[47]

## AUTHOR CONTRIBUTIONS

Conceptualization: Z.W., B.Q., W.T., J.J. Chemical synthesis: B.Q., F.L., F.R.F. On-surface synthesis and STM characterization: Z.W., P.H.J, J.L., X.L., M.F.C. DFT calculations: W.T., J.J., Z.W., B.Q., S.G.L. Funding acquisition and project supervision: S.G.L., F.R.F., M.F.C. The manuscript was written through contributions of all authors.

## ACKNOWLEDGEMENT

This work was primarily funded by the US Department of Energy (DOE), Office of Science, Basic Energy Sciences (BES), Materials Sciences and Engineering Division, under contract DE-AC02-05-CH11231 (Nanomachine program KC1203) (on-surface synthesis, STM imaging, DFT calculations) and contract DE-SC0023105 (molecular design). Support was also provided by the Office of Naval Research under awards N00014-19-1-2596 (image analysis) and N00014-24-1-2134 (molecular synthesis), and by the National Science Foundation under Grant No. CHE-2204252 (STM spectroscopy). Z.W. and B.Q. acknowledge support from Kavli ENSI Graduate Student Fellowships. F.R.F. acknowledges generous support by the Heising-Simons Faculty Fellows Program at UC Berkeley (development of synthetic facilities). We thank Dr. Hasan Çelik and the UC Berkeley NMR facility in the College of Chemistry (CoC-NMR) for assistance with spectroscopic characterization. Instruments in the CoC-NMR are supported in part by National Institutes of Health (NIH) award no. S10OD024998. We thank Drs. Dave Small, Cathleen Durkin, Azhagiya Singam, and the UC Berkeley Molecular Graphics and Computation Facility in the College of Chemistry (CoC-MGCF) for computational resources. The CoC-MGCF is supported in part by the National Institutes of Health (NIH) under award no. S10OD034382. We thank Dr. Zhongrui Zhou at the UC Berkeley QB3 Chemistry Mass Spectrometry Facility for assistance with mass spectroscopic characterization.

## FIGURES

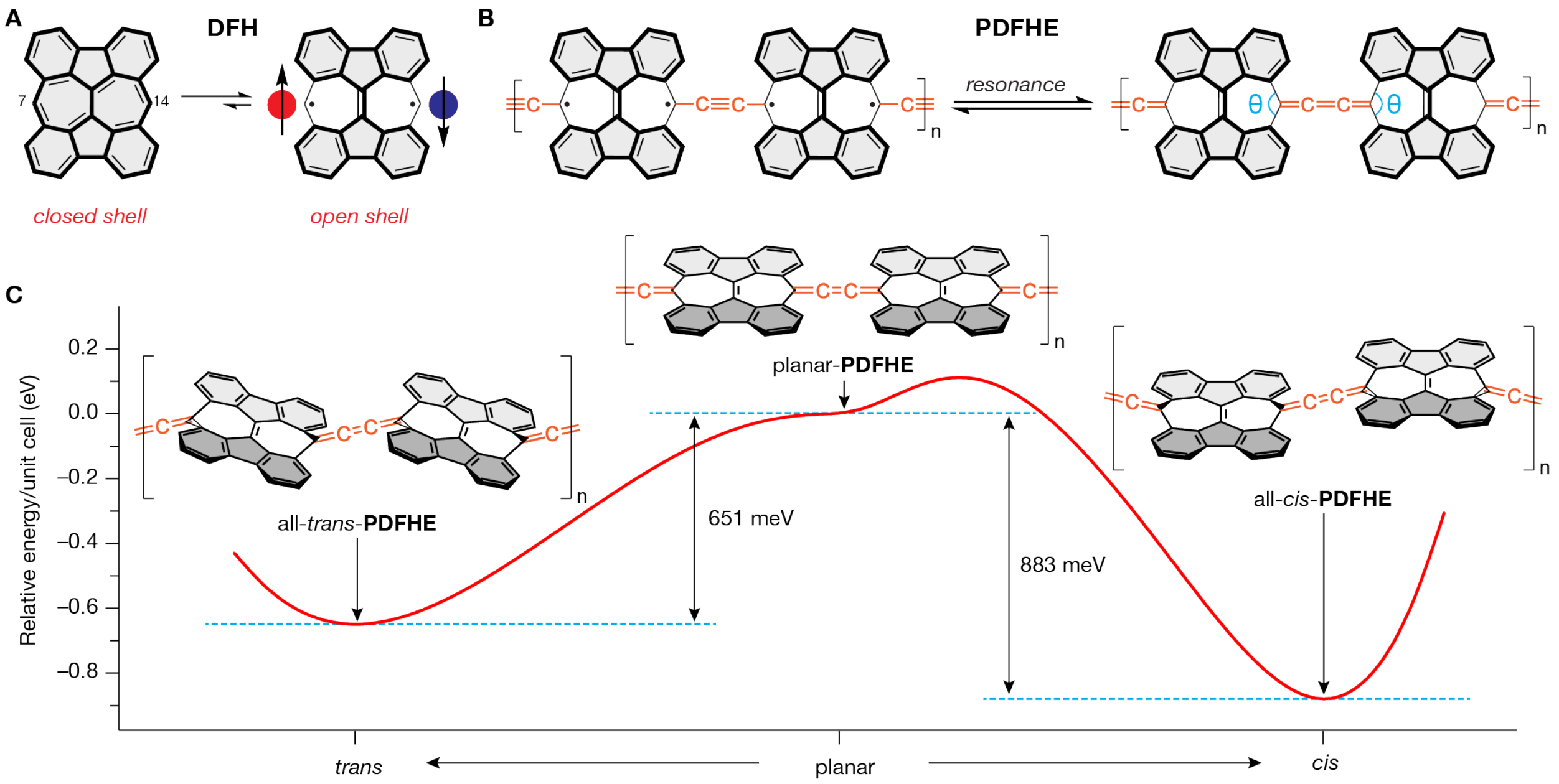

**Figure 1.** Structural distortion in PDFHE. (A) Schematic illustration of two possible electron configurations for the DFH core (closed shell vs. open shell). (B) Schematic representation of two resonance structures for the polymer PDFHE (open shell vs. closed shell cumulene). (C) DFT-computed energy diagram as a function of distortion coordinate shows the energetic stability of three primary structures: all-*trans*-PDFHE, planar-PDFHE, and all-*cis*-PDFHE. The planar structure occupies a higher-energy saddle point whereas the *trans* and *cis* structures are both at local energy minima.

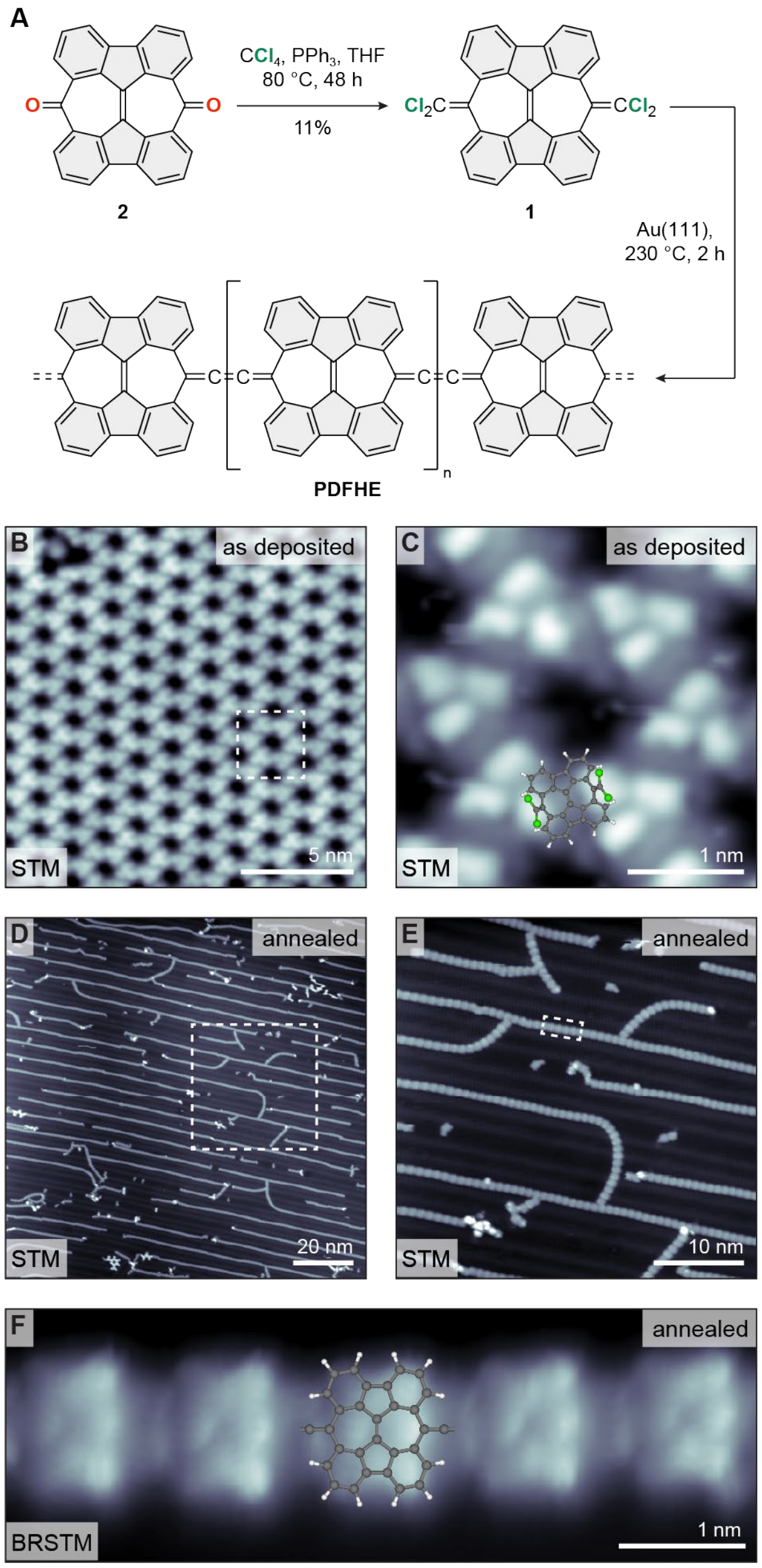

**Figure 2.** Bottom-up synthesis of PDFHE. (A) Schematic representation of the solution-based synthesis of molecular precursor **1**, and on-surface synthesis of PDFHE. (B) STM topographic image of self-assembled island of **1** on Au(111) (sample bias $V_S$ = –200 mV, tunneling current $I_t$ = 100 pA). (C) Close-up STM topographic image of precursors on Au(111) ($V_S$ = –100 mV, $I_t$ = 100 pA). (D) STM topographic image of PDFHE polymer chains on Au(111) after annealing to $T$ = 230 °C for 2 h ($V_S$ = –1600 mV, $I_t$ = 30 pA). (E) Close-up STM topographic image of PDFHE chains on Au(111) ($V_S$ = –1600 mV, $I_t$ = 30 pA). (F) Constant-height bond-resolved STM (BRSTM) image of a segment of PDFHE ($V_S$ = 5 mV, $V_{ac}$ = 30 mV, $f$ = 533 Hz, CO-functionalized tip). All STM experiments performed at $T$ = 4.7 K.

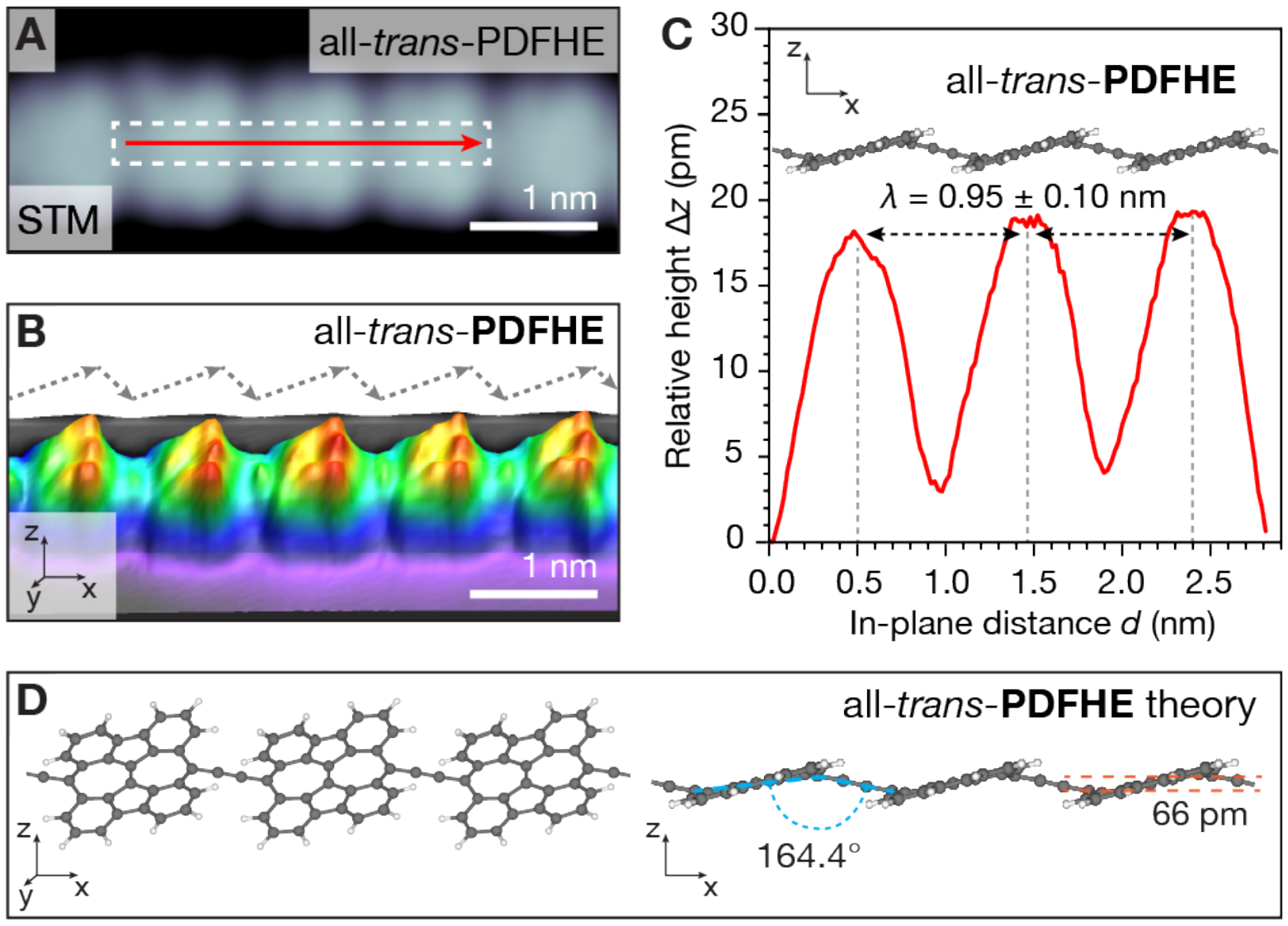

**Figure 3.** Structural distortion of all-*trans*-PDFHE. (A) STM topographic image of a segment of all-*trans*-PDFHE on Au(111) ($V_S$ = –400 mV, $I_t$ = 50 pA). The solid red arrow indicates the direction along which the height profile was extracted for (C). (B) 3D-rendered perspective of the BRSTM image in (Figure 2F). (C) Height line-cut extracted along the solid red arrow in (A). (D) Oblique (left) and side (right) views of the DFT-optimized structure of freestanding all-*trans*-PDFHE. All STM experiments performed at $T$ = 4.7 K.

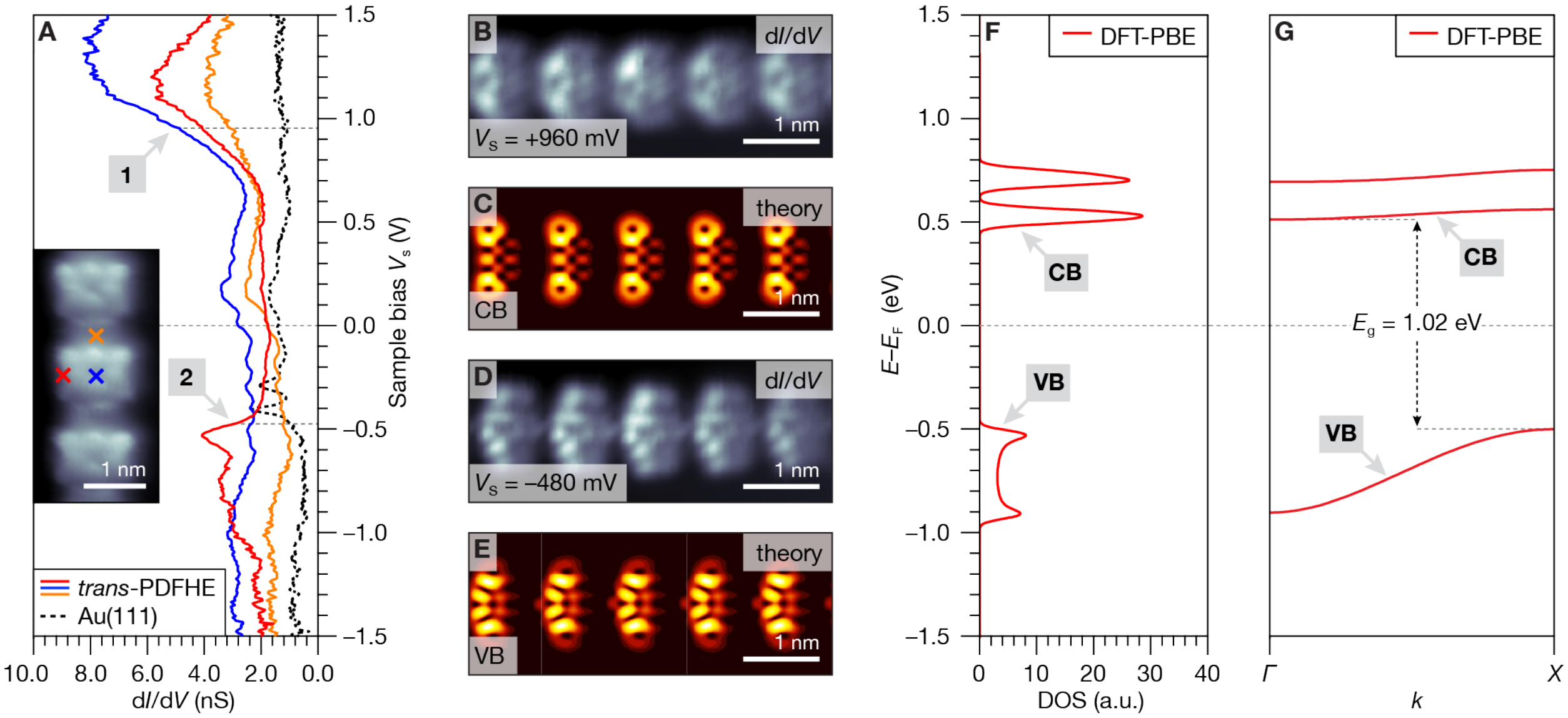

**Figure 4**. Electronic structure of all-*trans*-PDFHE. (A) d$I$/d$V$ point spectra acquired on all-*trans*-PDFHE on Au(111) at the locations indicated by colored crosses in the inset (inset: bond-resolved STM image of a PDFHE segment) ($V_{ac}$ = 4 mV, $f$ = 533 Hz, CO-functionalized tip). (B) Constant-height d$I$/d$V$ map of all-*trans*-PDFHE recorded at $V_S$ = +960 mV ($V_{ac}$ = 30 mV, $f$ = 533 Hz, CO-functionalized tip). (C) DFT-PBE calculated LDOS map at an energy corresponding to the conduction band-edge of freestanding all-*trans*-PDFHE. (D) Constant-height d$I$/d$V$ map of all-*trans*-PDFHE recorded at $V_S$ = –480 mV ($V_{ac}$ = 30 mV, $f$ = 533 Hz, CO-functionalized tip). (E) DFT-PBE calculated LDOS map at an energy corresponding to the valence band-edge of freestanding all-*trans*-PDFHE. (F) DFT-PBE calculated DOS of freestanding all-*trans*-PDFHE (spectrum broadened by a 27 meV width Gaussian). (G) DFT-PBE calculated band structure of freestanding all-*trans*-PDFHE. All STM experiments performed at $T$ = 4.7 K.